\documentclass[12pt]{article}
\usepackage{amssymb,amsmath,epsfig}

\begin{document}
\title{\bf Singularity in Gravitational Collapse of Plane Symmetric Charged Vaidya Spacetime}

\author{M. Sharif \thanks{msharif@math.pu.edu.pk} and Aisha
Siddiqa \thanks{asmth09@yahoo.com}\\
Department of Mathematics, University of the Punjab,\\
Quaid-e-Azam Campus, Lahore-54590, Pakistan.}

\date{}

\maketitle

\begin{abstract}
We study the final outcome of gravitational collapse resulting from
the plane symmetric charged Vaidya spacetime. Using the field
equations, we show that the weak energy condition is always
satisfied by collapsing fluid. It is found that the singularity
formed is naked. The strength of singularity is also investigated by
using Nolan's method. This turns out to be a strong curvature
singularity in Tipler's sense and hence provides a counter example
to the cosmic censorship hypothesis.
\end{abstract}


Oppenheimer and Snyder are the pioneers for the description of
gravitational collapse of stars \cite{1}. The study of gravitational
collapse is motivated by the fact that it represents one of the few
observable phenomena in the universe. The end state of a continual
gravitational collapse of a massive star is an important issue in
gravitation theory. According to Penrose, gravitational collapse of
a star gives rise to a spacetime singularity \cite{2} provided the
conditions such as trapped surface formation etc are satisfied.
Also, the singularity theorems of Hawking and Penrose provide a
strong reason to believe that a singularity occurs at the end of
gravitational collapse \cite{3}. The spacetime singularity is a
region where curvature and densities are infinite and their physical
description is not possible. These singularities are of two kinds:
naked if it is visible otherwise it is a clothed singularity, i.e.,
a black hole.

It would be interesting to investigate whether the singularity
forming at the end of gravitational collapse is observable. There
is an important conjecture related to the singularities known as
cosmic censorship hypothesis (CCH) given by Penrose \cite{5}. This
states that the collapse of a physically reasonable initial data
yields a spacetime singularity which is always hidden behind the
event horizon. It has two versions, i.e., weak and strong.
According to the weak version, singularity formed by gravitational
collapse is not visible to a far away observer. The strong cosmic
censorship hypothesis states that the singularity cannot be
observed even by an observer who is very close to it. Wald
\cite{6} discussed some examples to justify the validity of weak
form of CCH.

Despite of several attempts, there is no proof available for CCH and
it remained an open problem. However, significant progress has been
made in trying to find counter examples to CCH. Papapetrou \cite{7}
was the first who showed that the Vaidya solution \cite{8} could
give rise to naked singularities. This solution is widely used for
discussing counter examples to CCH. Using the concept of
gravitational lensing (GL), Virbhadra et al. \cite{9} introduced a
new tool for examining naked singularities. Gravitational lensing is
the process of bending of light around a massive object such as a
black hole. Virbhadra and Ellis \cite{10} discussed GL by the
Schwarzschild black hole. It was found that the relativistic images
guarantee the Schwarzschild geometry close to event horizon. The
same authors \cite{11} also analyzed GL by a naked singularity.
Claudel et al. \cite{12} proved that the necessary and sufficient
condition for the black hole to be surrounded by a photon sphere is
that a reasonable energy condition holds. Virbhadra and Keeton
\cite{13} showed that weak CCH can be examined observationally
without any uncertainty. Virbhadra \cite{14} found that Seifert's
conjecture is supported by the naked singularities forming during
Vaidya null dust collapse. The same author developed an improved
form of CCH using GL phenomenon \cite{15}.

Lemos \cite{16} showed that the gravitational collapse of a
spherical matter distribution in an anti-de Sitter spacetime form
naked singularities, violating CCH while for the cylindrical or
planar case, the collapse form black holes rather than naked
singularities supporting CCH. Ghosh \cite{17} introduced charged
null fluid in the results found by Lemos. Harko and Cheng \cite{18}
showed that a naked singularity is found in collapse of strange
quark matter with Vaidya geometry for a particular choice of
parameters. Sharif and his collaborators \cite{18a}-\cite{18f}
discussed gravitational collapse in a variety of papers and also the
effects of electromagnetic field are analysed.

Joshi and his collaborators \cite{20}-\cite{20a} used dust collapse
models to discuss physical features that caused the occurrence of
naked singularities. The results are generalized using type $I$
matter. They concluded that shearing forces and inhomogeneity within
the collapsing matter lead to the formation of naked singularities.
Zade et al. \cite{21} found that the singularity is naked in
monopole charged Vaidya spacetime.

In this brief paper, we analyze the singularity occurring in plane
symmetric charged Vaidya spacetime. For this purpose, the
spacetime for imploding radiations is given by \cite{22}
\begin{equation}\label{1}
ds^2=(\frac{2m(v)}{r}-\frac{e^2(v)}{r^2}) dv^2+2drdv+r^2(dx^2+dy^2).
\end{equation}
Here $-\infty<x,~y<\infty$ which describe 2-dimensional space and
has topology $R\times R,~-\infty<v<\infty$ is null coordinate,
called the advanced Eddington time and $0\leq r<\infty$ is the
radial coordinate. This represents solution of the Einstein-Maxwell
field equations with plane symmetry. We number the coordinates as
$x^0=v,~x^1=x,~x^2=y$ and $x^3=r$.

We take the matter as charged null dust for which the
energy-momentum tensor is \cite{17}
\begin{equation}\label{2}
T_{ab}={\rho}{\ell_{a}}{\ell_{b}}+T_{ab}^{em},
\end{equation}
where the part $\rho\ell_{a}\ell_{b}$ is for null dust, $\rho$ is
the energy density of the null dust and $\ell_{a}$ is a null
vector defined as
\begin{equation}\label{3}
\ell_{a}={\delta^{0}_{a}},\quad{\ell_{a}\ell^{a}=0}.
\end{equation}
The electromagnetic energy-momentum tensor is
\begin{equation}\label{4}
T_{ab}^{em}=\frac{1}{4\pi}[F_{ac}F_{b}^{c}-\frac{1}{4}g_{ab}F_{cd}F^{cd}],
\end{equation}
where $F_{ab}$ is the electromagnetic field tensor given by
\begin{equation}\label{5}
F_{ab}=\frac{e(v)}{r^2}(\delta^{0}_{a}\delta^{3}_{b}-\delta^{0}_{b}\delta^{3}_{a}).
\end{equation}
Using the Einstein field equations, $G_{ab}=\kappa T_{ab}$, we
obtain
\begin{equation}\label{7}
\rho=\frac{1}{4\pi r^3}(r\dot{m}-e\dot{e}),
\end{equation}
where dot denotes differentiation with respect to $v$. The weak
energy condition is satisfied if $\rho\geq0$, which is possible
only when $r\dot{m}-e\dot{e}\geq0$, i.e., $r\geq
\frac{e\dot{e}}{\dot{m}}$. We know from the literature \cite{23}
that Lorentz force prevents the particle to move into the region
where $r< \frac{e\dot{e}}{\dot{m}}$, so the energy condition is
always satisfied.

It is assumed that the first wavefront arrives at $r=0$ at time
$v=0$ and the final wavefront reaches the center at $v=T$. A
singularity of growing mass develops here at $r=0$. For $v<0$, the
spacetime is Minkowski with $m(v)=e(v)=0$ and for $v>T$, the
spacetime settles to plane symmetric Reissner Nordstr$\ddot{o}$m
solution. Thus, for $v=0$ to $v=T$, the spacetime is plane symmetric
charged Vaidya spacetime.

In order to check whether the singularity is naked or clothed, we
examine the behavior of radial null geodesic. This is defined as
\cite{24}
\begin{equation}\label{8}
ds^2=0,\quad  dx=0=dy.
\end{equation}
If the radial null geodesic equation admits at least one real and
positive root then the singularity is naked. The geodesic equations
for charged Vaidya spacetime, using Eq.(\ref{8}), are given by
\begin{equation}\label{9}
\ddot{v}+(\frac{mr-e^{2}}{r^{3}})\dot{v}^{2}=0,
\end{equation}
\begin{equation}\label{10}
\ddot{r}+[\frac{\dot{m}r-e\dot{e}}{r^{2}}-\frac{(2mr-e^{2})(mr-e^{2})}{r^{5}}]\dot{v}^{2}
-2(\frac{2mr-e^{2}}{r^{3}})\dot{r}\dot{v}=0.
\end{equation}
Now $m(v)$ and $e(v)$ are unknown functions of $v$. For analytic
solution, we take
\begin{equation}\label{11}
m(v)=\frac{\lambda v}{2},\quad e^2(v)=\mu v^2;\quad \lambda,~\mu>0.
\end{equation}
Using these values in Eq.(\ref{1}), it follows that
\begin{equation}\label{12}
ds^{2}=(\frac{\lambda v}{r}-\frac{\mu
v^2}{r^2})dv^{2}+2drdv+r^2(dx^2+dy^2).
\end{equation}
This represents a self-similar metric and $X=\frac{v}{r}$ is the
self-similarity variable.

The radial null geodesics for this spacetime are given by
\begin{equation}\label{13}
\frac{dv}{dr}=\frac{2}{\frac{-\lambda v}{r}+\frac{\mu v^2}{r^2}}
\end{equation}
which has a singularity at $r=0,~v=0$. Suppose that $X_{0}$ is the
limiting value of $X$ as the singularity is approached, i.e.,
\begin{equation}\label{14}
X_{0}=\lim_{r\rightarrow 0,v\rightarrow 0}X=\lim_{r\rightarrow
0,v\rightarrow 0}\frac{v}{r}=\lim_{r\rightarrow 0,v\rightarrow
0}\frac{dv}{dr}.
\end{equation}
Using Eq.(\ref{13}), this equation becomes
\begin{equation}\label{15}
\mu X_{0}^3-\lambda X_{0}^2-2=0.
\end{equation}
We know that \cite{25} \emph{every equation of odd degree has at
least one real root whose sign is opposite to that of its last term,
the coefficient of first term being positive}. This implies that
Eq.(\ref{15}) has at least one real and positive root independent of
the values of parameters $\lambda$ and $\mu$. Hence the singularity
occurring in plane symmetric charged Vaidya spacetime is at least
locally naked, in particular, for $\lambda=0.1,~\mu=0.1,~
X_{0}=3.09198$. Singularity arising in uncharged case can be
examined by taking $\mu=0$ in Eq.(\ref{15}). In this case, it turns
out that Eq.(\ref{15}) has no real root implying that the collapse
ends at a black plane as final state.

To see whether it is a scalar polynomial singularity or not, we find
the non-vanishing components of the Riemann tensor
\begin{eqnarray*}
R_{0113}&=&-(\frac{-mr+e^2}{r^2})=R_{0223},\\
R_{1212}&=&2mr-e^2,\quad R_{0303}=\frac{-2mr+3e^2}{r^4},\\
R_{0101}&=&-\frac{1}{r^4}(-r^4\dot{m}+r^3e\dot{e}+2m^2r^2-3mre^2+e^4).
\end{eqnarray*}
Using these values, the Kretschman scalar
$\mathcal{R}=R_{abcd}R^{abcd}$, becomes
\begin{equation}\label{18}
\mathcal{R}=\frac{4}{r^6}(4m^2+\frac{e^4}{r^2}-\frac{4me^2}{r}).
\end{equation}
Also in view of Eq.(\ref{11}) and $X=\frac{v}{r}$, it takes the form
\begin{equation}\label{19}
\mathcal{R}=\frac{4}{r^4}(\lambda^2X^2+\mu^2X^4-2\lambda\mu X^3).
\end{equation}
This shows that the Kretschman scalar diverges at the singularity
and hence singularity is a scalar polynomial.

Now we explore the strength of singularity using Nolan's method
\cite{26} according to which a singularity is strong at $r=0$ if
$\dot{r}$ has zero or infinite limit along every causal geodesic
approaching the singularity. The strength of singularity is
important in the sense that CCH may not be ruled out for weak
naked singularities. We assume that $\dot{r}$ is non-zero and
finite as the singularity is approached, i.e.,
\begin{equation}\label{17}
\dot{r}\sim h_{0}\Rightarrow r\sim kh_{0}
\end{equation}
and
\begin{equation}\label{18}
\frac{\dot{v}}{\dot{r}}=X_{0}\Rightarrow v\sim kh_{0}X_{0},
\end{equation}
here dot denotes differentiation with respect to affine parameter
$k$. Using Eqs.(\ref{17}) and (\ref{18}), Eq.(\ref{9}) implies that
\begin{equation}\label{19}
\ddot{v}=Ck^{-1},\quad
C=-(\frac{\lambda}{2}-\mu X_{0})h_{0}X_{0}^{3}.
\end{equation}

Since $k=0$ as the singularity is approached which implies that
$\ddot{v}$ is undefined  and hence $\dot{v}$ is undefined. However,
we already have $\dot{v}=h_{0}X_{0}$. These values are consistent
only if $C=0$ giving rise to
\begin{equation}\label{21}
X_{0}=\frac{\lambda}{2\mu}.
\end{equation}
Replacing this value of $X_{0}$ in Eq.(\ref{15}), we get
\begin{equation}\label{22}
\lambda^3+16\mu^2=0
\end{equation}
which is not possible as $\lambda$ and $\mu$ are positive. Thus
the inconsistency of Eqs.(\ref{18}) and (\ref{19}) imply that our
supposition is wrong, i.e., $\dot{r}$ is either zero or infinite.
Hence singularity is strong in Tipler \cite{27} sense providing a
counter example to weak form of CCH.

Finally, we conclude that the singularity is at least locally
naked violating weak CCH. The existence of positive roots of
Eq.(\ref{15}) only show that the null geodesics are coming out
from the singularity but nothing could be said about the escape of
these geodesics from the boundary of the collapsing matter.
However, the mass function can be chosen in such a way that a
locally naked singularity becomes globally naked \cite{28}.

\end{document}